# Disentangling the Behavioural Variability of Confined Cell Migration

David B. Brückner[1], Alexandra Fink[2], Joachim O. Rädler[2], Chase P. Broedersz[1,#]

[1] Arnold-Sommerfeld-Center for Theoretical Physics and Center for NanoScience, Ludwig-Maximilians-Universität, München
[2] Faculty of Physics and Center for NanoScience, Ludwig-Maximilians-Universität, München

[#] corresponding author (c.broedersz@lmu.de)

Cell-to-cell variability is inherent to numerous biological processes, including cell migration. Quantifying and characterizing the variability of migrating cells is challenging, as it requires monitoring many cells for long time windows under identical conditions. Here, we observe the migration of single human breast cancer cells (MDA-MB-231) in confining two-state micropatterns. To describe the stochastic dynamics of this confined migration, we employ a dynamical systems approach. We identify statistics to measure the behavioural variance of the migration, which significantly exceed those predicted by a population-averaged stochastic model. This additional variance can be explained by the combination of an 'aging' process and population heterogeneity. To quantify population heterogeneity, we decompose the cells into subpopulations of slow and fast cells, revealing the presence of distinct classes of dynamical systems describing the migration, ranging from bistable to limit cycle behaviour. Our findings highlight the breadth of migration behaviours present in cell populations.

**Introduction**

On all scales of life, the behaviour of single organisms is intrinsically variable, both as a function of time and between individuals. A number of studies have quantified the inter-individual variations in behaviour in a broad range of systems, from bacteria [1–4] to protozoans [5,6], fruit flies [7], mice [8], and humans [9,10]. However, even in populations of single mammalian cells with identical genomes, the intrinsic stochasticity of intra-cellular processes such as gene expression, cytoskeletal rearrangement and protein localization can lead to large differences in the proteomes of individual cells [11–15]. Further downstream, this diversity leads to cell-to-cell variability (CCV), i.e. phenotypic or population heterogeneity, at the level of whole-cell behaviours, such as growth rate, drug response, morphology, and migration [16–20]. In the case of migrating cells, CCV has profound implications at larger scales for the behaviour of cell clusters, such as their ability to chemotax [21], invade surrounding tissue [22] and perform collective motion [23,24]. Thus, CCV is increasingly well understood at the molecular level, and its implications for collective behaviour are becoming clearer. However, the detection and quantitative characterization of this variability in ensembles of individual migrating cells remains challenging; such a characterization requires both an appropriate theoretical framework and data sets where migrating cells are monitored over a sufficiently long time under identical conditions.

Cell migration is a key feature of many cell types, including immune, epithelial and cancer cells. The movement of such cells is powered by an intricate machinery that relies on the coordination of vast numbers of molecular constituents, whose collective dynamics are key in generating the persistent motion of cells [25,26]. Despite this underlying complexity, a number of studies have shown that the emergent migratory dynamics at the cellular scale can be quantitatively captured by relatively simple stochastic equations of motion [6,27–32]. This approach has been successfully applied to migration on uniform, two-dimensional surfaces [6,27–30,33], and recently to cells migrating in confinement [32]. However, to determine the structure and parameters of such stochastic models, the dynamics are typically averaged across different cells and over time, yielding ensemble- and time-averaged (ETA) models that describe the average member of a cell population, and thus fail to capture phenotypic heterogeneity [15,34]. Similarly, bottom-up models for cell motility typically assume that all cells in a population can be described by a common set of parameters [35–42].

To characterize heterogeneity in cell migration, a super-statistical framework was previously applied to cell migration [43]. By fitting a persistent random motion model with time-dependent parameters to migration in a variety of environments, large variations in cell behaviour were revealed both in time and between individuals. Another approach has been to construct a phenotypic space through dimensional reduction, which identifies the aspects of behaviour that vary most between individuals [5]. However, in both approaches, the presence of time- and population-heterogeneity is assumed from the start. It thus remains unclear how much of the observed variability is due to real heterogeneity, and how much is due to the intrinsic stochasticity of the migration process. A central difficulty in detecting phenotypic heterogeneity in cell migration is the short time window of a trajectory of an individual cell, which is limited by cell division. Thus, even for a hypothetical population of identical cells whose migration can be described by a single equation of motion, one would expect to observe statistical differences between individual trajectories due to the limited observation time of the experiment. A key challenge is therefore to distinguish such apparent variability in short trajectories from real inter-individual variability. To detect and characterize CCV in behaviour, it would therefore be helpful to have a model system for which a quantitative theory for the ensemble-averaged behaviour of the cells has been developed. Such an ETA model can then provide a reference to identify observables that are sensitive to potential phenotypic heterogeneity in the population.

In previous work [32], we analyzed the time-trajectories of single breast cancer cells (MDA-MB-231) confined to two-state micropatterns, in which these cells repeatedly transit across a thin constriction. This system mimics the extra-cellular environment cells face in physiological settings, in which they navigate confining structures and squeeze through thin channels and constrictions [44,45,54–56,46–53]. An important advantage of this setup is that it allows us to quantitatively monitor the behaviour of a large population of cells in identical, standardized surroundings, suppressing extrinsic sources of heterogeneity in the trajectories. The overall dynamics of this two-state migration are well described by a stochastic equation of cell motion, with a deterministic and stochastic contribution that can be inferred from the short time-scale data observed experimentally. This model accurately reproduces the ensemble- and time-averaged statistics including the dwell time distribution, the distributions of position and velocity, and the velocity autocorrelation function.

Here, we develop a theoretical approach to address the behavioural variability of confined cell migration arising through phenotypic heterogeneity and time-dependent dynamics. To this end, we identify a statistic that reveals the intrinsic variability of the migration in two-state micropatterns: the variance in the number of transitions a single cell performs in a given time. The sensitivity of this statistic to variability arises because it measures single-cell behaviour accumulated over long periods of time, beyond the time-scale of a single transition, and compares this to the population average. Indeed, we find that this variance significantly supersedes the prediction of the ensemble-averaged model. We account for this observation by two effects: a time-dependent 'aging' effect and phenotypic heterogeneity. From the trajectory data, we show that the average acceleration of the cells gradually decreases over time, leading to a decrease in hopping activity. Furthermore, we propose to capture phenotypic heterogeneity by inferring migration behaviour for subpopulations of cells. Specifically, we find that by splitting the cell population by transition time into only two subpopulations, we can capture most of the inter-cell variance. These subpopulations are described by distinct classes of dynamical systems, with fast cells performing limit cycle oscillations while slow cells exhibit excitable bistability. Thus, our approach reveals a striking breadth of migratory behaviours within a cell population, paving the way to investigate the origins and functional implications of such variability.

# Results

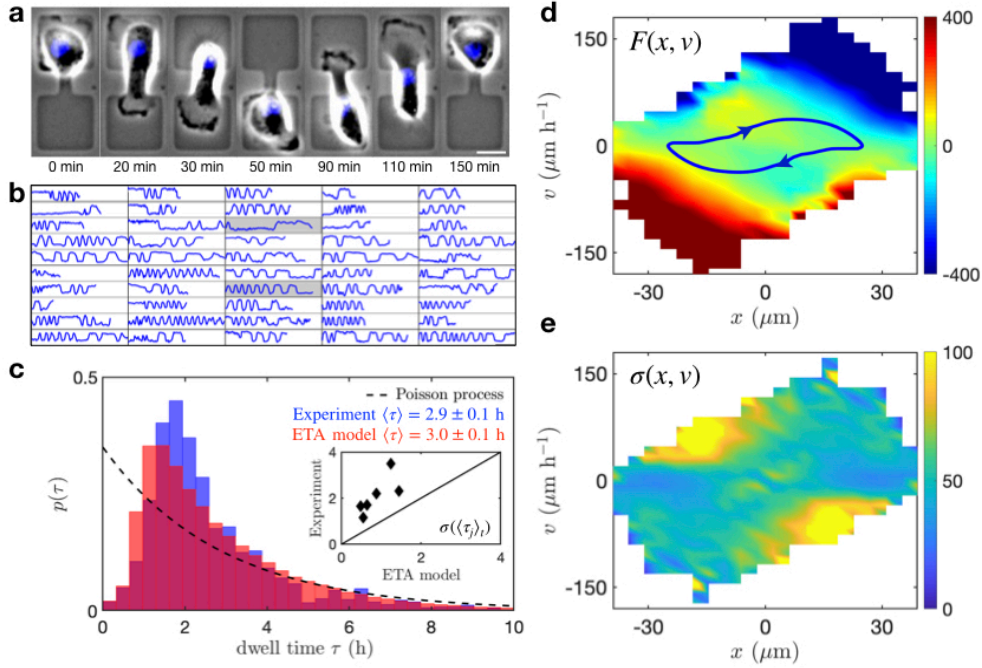

**Fig 1. Equation of motion for an ensemble of cells migrating on two-state micropatterns ($L$ = 16 μm).** (a) Time series of a single cell migrating on a two-state micropattern with bridge length $L$ = 16 μm (scale bar: 20 μm). (b) Trajectories of the cell nucleus as a function of time, plotted from $t$ = 0 to $t$ = 50 h, for a collection of cells. The two trajectories shaded in grey correspond to supplementary videos S1 and S2, providing examples of a 'slow' and 'fast' cell, respectively. (c) Distribution of dwell times observed experimentally (blue) and predicted by an ensemble-averaged equation of motion (red). The dashed line indicates the probability distribution of a Poisson process with the same mean dwell time, defined by $p(\tau) = \upsilon e^{-\upsilon \tau}$, with $\upsilon = 1/\langle \tau \rangle_{t,e}$. Inset: inter-cellular variance $\sigma(\langle \tau(j) \rangle_t)$ observed in experiment plotted against the prediction by the ETA model. (d) The deterministic term $F(x,v)$ of the ETA model in units of μm/h². (e) The ETA model noise strength $\sigma(x,v)$ in units of μm/h$^{3/2}$.

**Cells perform non-Poissonian transitions on two-state micropatterns**

To develop a quantitative framework for the variability of migrating cells, we employ two-state micropatterns, in which cells are confined to a standardized geometry consisting of two square adhesive sites connected by a bridge of length $L$ (Fig. 1a). In this environment, we previously found that MDA-MB-231 breast cancer cells perform repeated transitions between the adhesive sites [32]. This setup provides a minimal system to study how cells respond to the presence of thin constrictions and ensures that all cells in the population interact with an identical micro-environment.

Interestingly, the trajectories of the cell, which we define by the position of the nucleus center, exhibit large variations both in time and across individuals in the population (Fig. 1b; supplementary movies S1 and S2). Accordingly, the probability distribution of dwell times exhibits a large spread (Fig. 1c). This raises a key question: Can the experimentally observed variation be explained by a single stationary stochastic process with time-independent parameters identical for all individuals?

The two-state migration assay provides an ideal platform to address this question. Previously, we have identified a stochastic equation of cell motion that captures the ensemble- and time-averaged behaviour of cells on this micropattern [32]. This equation takes the form

$$\frac{dv}{dt} = F(x,v) + \sigma(x,v)\eta(t) \qquad (1)$$

where $\eta(t)$ represents Gaussian white noise with $\langle\eta(t)\rangle = 0$ and $\langle\eta(t)\eta(t')\rangle = \delta(t-t')$. Here, the deterministic term $F(x,v) = \langle\dot{v}|x,v\rangle_{t,e}$ (Fig. 1d) and the noise strength $\sigma^2(x,v) = \Delta t\langle[\dot{v} - F(x,v)]^2|x,v\rangle_{t,e}$ (Fig. 1e) are inferred from the short time-scale cell dynamics measured experimentally [57–59], where $\langle\Theta\rangle_{t,e}$ indicates time- and ensemble-averaging of the observable $\Theta$ (see Methods). The terms $F(x,v)$ and $\sigma(x,v)$ thus show the average acceleration of the cell as a function of its position in $xv$-phase space, and the magnitude of stochastic fluctuations, respectively. In the case of the MDA-MB-231 cell line considered here, the deterministic term exhibits non-linear dynamics in the form of a limit cycle, indicating that the cellular transitions can be interpreted as noisy non-linear oscillations (Fig. 1d). We have shown that this model accurately captures the key statistics of the cell migration, including the dwell time distribution (Fig. 1c), as well as the distributions of position and velocity and the velocity correlation function [32].

To probe the limits of the ETA approach, we first test whether it also quantitatively predicts the dwell time distributions at the single-cell level. To do so, we simulate the model with the same observation time per cell as recorded experimentally. To measure cell-to-cell variability, we obtain the average dwell time $\langle\tau(j)\rangle_t$ for each single cell $j$ and then determine the standard deviation $\sigma(\langle\tau(j)\rangle_t)$ of these single-cell averages, giving a measure of the inter-cell variance. Clearly, in the limit of infinitely long trajectories, the inter-cell variance in the ensemble-averaged model should vanish; therefore, the finite variance we find here for simulated trajectories is simply due to finite trajectory length. Interestingly, however, we find that the experimental inter-cell variance significantly exceeds that predicted by the ETA model for this trajectory length (inset Fig. 1c). This suggests that there are significant variations from cell to cell, which cannot be accounted for by the ETA model. We therefore hypothesize that the cell population exhibits phenotypic heterogeneity detectable at the level of cell trajectories.

The presence of phenotypic heterogeneity could lead to apparent temporal correlations in behaviour that persist for the whole life of a cell, up to division. To further quantify the overall variability in the transition behaviour, we thus need to consider statistics that could be sensitive to correlations over a range of time-scales spanning from a single transition time up to the division time. A common way to quantify the overall variability of transition processes is to determine the ratio between mean and variance in the number of transitions within a given time interval. Thus, we measure the distribution function of the hopping statistics, which is given by the probability distribution $p(N,T)$ to observe $N$ transitions of a cell in a given time interval $T$ (Fig. 2a). Importantly, to determine the statistics of the observable $N$, we monitor the behaviour of a single cell over a wide range of time-scales, extending well beyond the average transition time. We therefore expect this statistic to be particularly sensitive to behavioural heterogeneity in a population, where some cells might have the tendency to consistently perform slower- or faster-than-average transitions.

As the observation time interval is increased, both the mean $\mu(T)$ and variance $\Sigma^2(T)$ of these distributions increase (inset Fig. 2a). The simplest 'hopping' process with such a distribution is the Poisson process, for which $\mu = \Sigma^2$. Interestingly, this is not the case here (Fig. 2b). Instead, at short time-scales, the transitions are sub-Poissonian with a mean that exceeds the variance ($\mu > \Sigma^2$), while beyond a cross-over time-scale $T^*$, the variance exceeds the mean, $\mu < \Sigma^2$, indicating a super-Poissonian regime. In fact, as we vary the length $L$ of the connecting bridge, this behaviour is qualitatively reproduced in all cases (Fig. 2b). The cross-over time-scale $T^*$ scales linearly with the average dwell time $\langle \tau \rangle_{t,e}$, and typically exceeds it by a factor of approximately 4 (inset Fig. 2b).

These results show that the ratio between mean and variance in this transition process exhibits a subtle dependence on the time-scale on which the system is observed. At short time-scales, up to a few times the average dwell time, we observe a smaller variance than in a Poisson process, which is defined by a transition probability that is constant in time. Similar sub-Poissonian behaviour is found, for instance, in the stepping behaviour of molecular motors, where it was rationalized as a consequence of an underlying multi-step scheme that has to be executed before a large step can be made [60]. Analogously, here we might argue that during the cellular transitions in the two-state confinement, a number of cytoskeletal and shape rearrangements have to occur between transitions, which leads to fewer short dwell times than would be expected from a Poisson process (Fig. 1c). On long time-scales, however, the transition variance starts to significantly exceed the mean. Such super-Poissonian statistics have previously been observed, e.g. in RNA-transcription [61], and were attributed to cell-to-cell variations in the transcription rates [62].

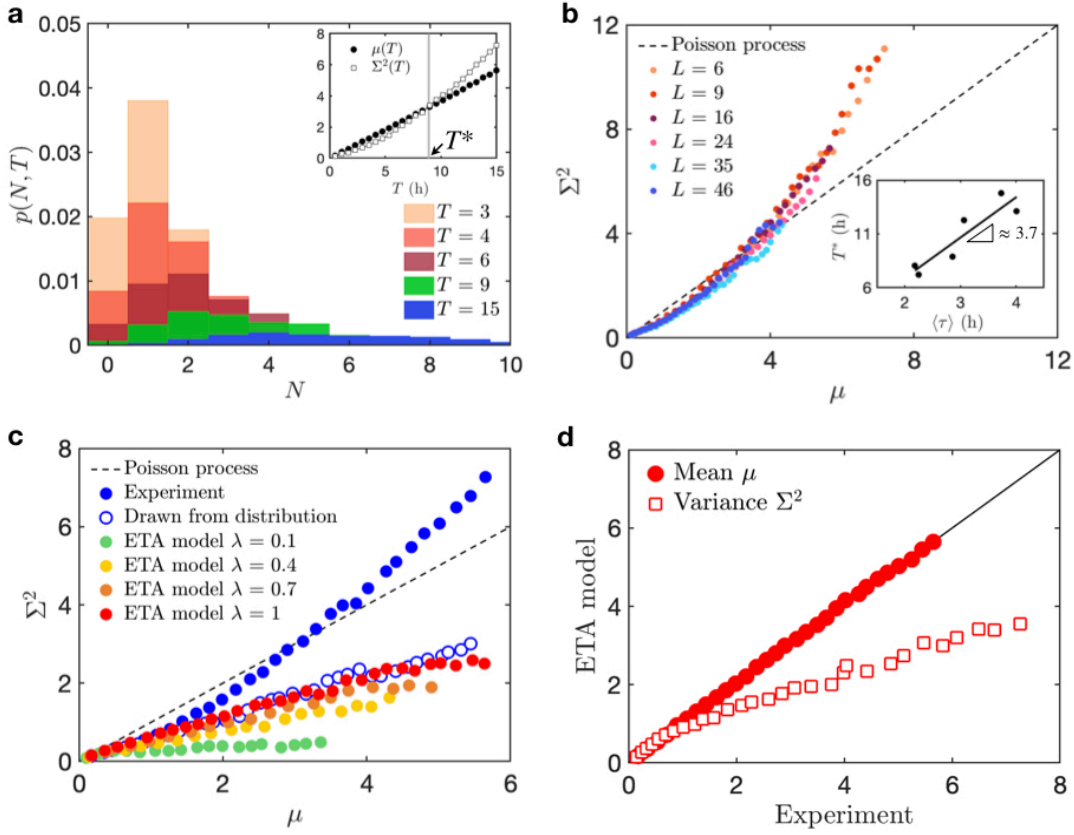

**Fig 2. Transition number statistics of hopping cells.** (a) Probability distributions of the number of transitions $N$ of a single cell in a given time interval $T$, indicated by the colors. Inset: Variance $\Sigma^2$ and mean $\mu$ of the number of transitions as a function of the time interval $T$. The grey vertical bar indicates the cross-over time-scale $T^*$. (b) Variance plotted against the mean for all bridge lengths. The dashed line corresponds to the Poisson behaviour $\mu = \Sigma^2$. Inset: Cross-over time-scale $T^*$ plotted as a function of $\langle \tau \rangle_{t,e}$. Black line corresponds to a linear fit. (c) Mean-variance curve observed experimentally (blue) compared to the prediction by the stochastic model (red) and the expectation from random sampling of the experimental dwell time distribution (open symbols). In green, orange and yellow, we show predictions of the model with a noise amplitude scaled by factors $\lambda = 0.1, 0.4, 0.7$. (d) Mean and variance predicted by the model plotted against experimental values. Panels (a), (c), (d) correspond to bridge length $L$ = 16 μm.

**Observed variability exceeds that of a single stochastic process**

The observation of non-Poissonian statistics does not necessarily indicate cell-to-cell variability. To investigate the origin of the non-Poissonian statistics in the hopping behaviour of cells, we test our inferred ETA model (Eq. (1)) by comparing the predicted mean-variance curve to that measured experimentally. Strikingly, the ETA model predicts a qualitatively different, purely sub-Poissonian trend (Fig. 2c). While this model accurately predicts the average dwell time $\langle \tau \rangle_{t,e}$, and thus also the mean of the hopping distribution $\mu$, it underestimates the variance (Fig. 2d).

The behaviour of the statistics predicted by the ETA model can be understood intuitively: in a system with no noise ($\sigma(x,v) = 0$), we observe regular limit cycle oscillations, with zero variance. To investigate how different noise levels theoretically affect the hopping variance, we introduce an artificial prefactor $\lambda$ to scale the noise: $\dot{v} = F(x,v) + \lambda\sigma(x,v)\eta(t)$. As $\lambda$ is increased from 0 to 1, we find that the variance gradually increases (Fig. 2c). At the experimentally inferred noise level $\lambda = 1$, the system still appears to be dominated by the underlying deterministic oscillation period, giving rise to a sub-Poissonian process at all time-scales. Thus, while the ETA model predicts many statistics of the motion correctly, including the short time-scale variability, it falls short of capturing variability in the hopping behaviour on longer time-scales.

Next, we investigate the origin of the super-Poissonian regime directly from experimental data. If we assume that all cells follow the same equation of motion with identical parameters, and that these parameters do not vary in time, then each cell can be described by the same time-independent dwell time distribution $p(\tau)$. In other words, the hopping distribution $p(N,T)$ should be fully constrained by $p(\tau)$ and the stationary correlations between hops [60]. Interestingly, if we generate a trajectory of 'hops' by random independent sampling from the population dwell time distribution, we recover the same sub-Poissonian mean-variance relation as predicted by the inferred ETA model (Fig. 2c). This indicates that the transition time of subsequent hops in the ETA model have negligible correlations, while the experimental data may exhibit correlations extending over multiple hops.

By time- and ensemble-averaging, we miss this aspect that influences the long time-scale variability of the motion. In the next sections, we therefore carefully assess the validity of our assumptions of time-invariance and population homogeneity.

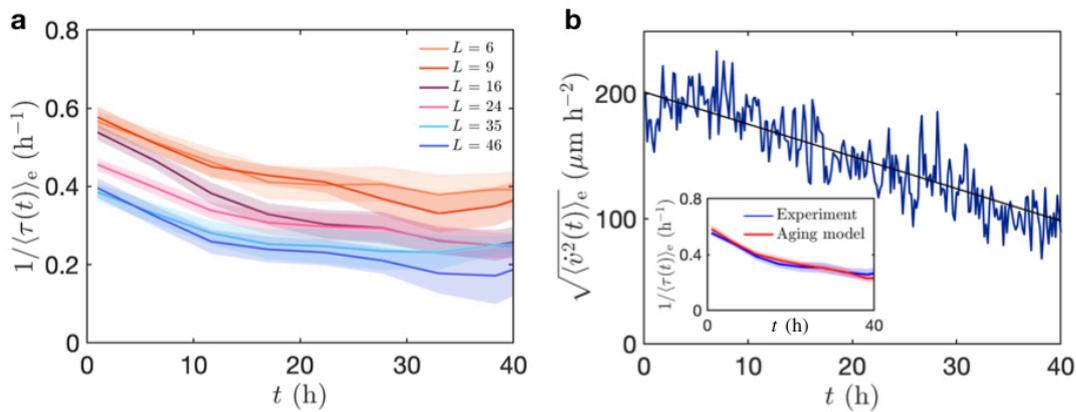

**Fig 3. Testing time-invariance.** (a) Hopping rates $1/\langle\tau(t)\rangle_e$ as a function of time for all bridge lengths. The quantity $\langle\tau(t)\rangle_e$ is averaged over the cell population and in time windows of width 5 h. Shaded intervals correspond to standard error in the mean. (b) Root-mean-square acceleration of the cells as a function of time ($L$ = 16 μm). Black line corresponds to a linear fit. Inset: Hopping rates as a function of time predicted by the aging model compared to the experimental observation ($L$ = 16 μm).

**Attenuation of cell dynamics enhances transition variability**

To investigate if the cellular hopping process indeed violates time-invariance, we measure the hopping rates $1/\langle\tau(t)\rangle_\text{e}$ as a function of time. The rate slowly decreases with time, indicating that the cell migration is slowing down over time (Fig. 3a). Similarly, the root-mean-square acceleration decreases roughly linearly as a function of time, by almost a factor of two (Fig. 3b). Note, however, that the cell population in this experiment is not cell-cycle-synchronized, such that *t* = 0 does not necessarily correspond to the same moment in the cell cycle for all cells. The decrease in acceleration we identify thus simply indicates the approximate trend observed in this experiment, giving the average behaviour of an aging population.

Clearly, the average acceleration is changing significantly over time. To incorporate this into the model, we postulate that most of the time-dependence can be accounted for by introducing an overall time-dependent pre-factor into the equation of motion, which we term the aging model:

$$\frac{dv}{dt} = \alpha(t)[F(x,v) + \sigma(x,v)\eta(t)] \qquad (2)$$

where $\alpha(t) = \sqrt{\langle\dot{v}^2(t)\rangle_\text{e}}/\sqrt{\langle\dot{v}^2\rangle_{t,\text{e}}}$, and $\sqrt{\langle\dot{v}^2\rangle_{t,\text{e}}}$ is the root-mean-square acceleration averaged over all time and cells. This implementation accurately reproduces the time-dependence of the hopping rates (inset Fig. 3b). Remarkably, this simple approach also significantly improves our estimate of the hopping variance $\Sigma^2$ (Fig. 4e). An alternative approach to incorporate time-dependence into our model would be to infer, for example, separate early- and late-stage models, corresponding to the first and second half of the experiment. We tested such an approach and found that the resulting predictions did not improve the estimate of the transition variability (Supplementary Section S1). However, even the more successful pre-factor aging approach underestimates the experimentally observed variance, suggesting that another effect needs to be accounted for.

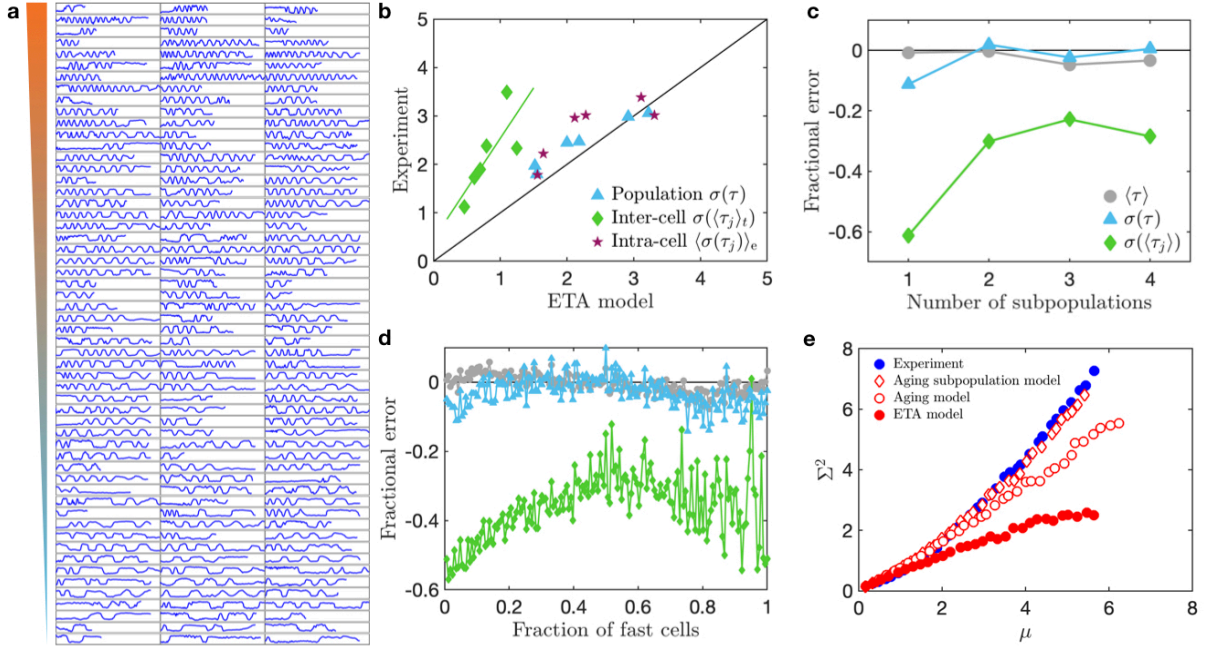

**Fig 4. Detection and characterization of cell-to-cell variability.** (a) All cell trajectories $x(t)$ plotted as a function of time, from $t$ = 0 to $t$ = 50 h. The cells are ranked by their average single-cell dwell time $\langle \tau(j) \rangle_t$, with the fastest cells at the top and the slowest at the bottom. Note, we exclude the first and last stay of each trajectory from the analysis throughout, as they are defined by the start and end of measurement, respectively, or by cell division, and are therefore not part of the hopping process. (b) Experimental observations plotted against ETA model predictions for various variance measures for all six bridge lengths: the population variance of all dwell times $\sigma(\tau)$, the inter-cell variance $\sigma(\langle \tau(j) \rangle_t)$, and the intra-cell variance $\langle \sigma(\tau(j)) \rangle_e$. All variances are plotted as standard deviations. The green line indicates a linear fit to the inter-cell variance points, which has a slope of 2. (c) Fractional error in the prediction of various quantities as a function of the number of subpopulation models. In each case, the population is divided evenly into subpopulations of equal size. (d) Fractional error in various quantities as a function of the location of the subpopulation split, defined as the number of fast cells in the two-type subpopulation model. Symbols are defined in the legend of panel (c). (e) Mean-variance curves of the ETA model (Eq. (1); red dots), the aging model (Eq. (2); open dots) and the aging subpopulation model (open diamonds) in which two subpopulation models are combined with the time-dependent prefactor. Experimental data is shown by blue dots. Panels (a), (c)-(e) correspond to bridge length $L$ = 16 μm.

**Population heterogeneity is captured by a decomposition into subpopulations**

In the previous section, we showed that by time averaging, we miss an important contribution to the hopping variance caused by the gradual decrease in the acceleration of the cells. Similarly, in the inference of the ETA model, we have employed ensemble averaging, and have thus assumed that all cells follow the same equation of motion with identical parameters. However, individual cells of the same cell line may have variations in their genome [63], but even in an isogenic population of cells, we expect there to be phenotypic differences between cells [11,12,17,64]. Thus, we next consider the impact of phenotypic heterogeneity on our model.

The ETA model performs well in predicting the variance of the population distribution $\sigma(\tau)$ (Fig. 4b). Moreover, the ETA model gives a good estimate of the intra-cellular variance, defined as the average variance of dwell times within the life time of a single cell $\langle\sigma(\tau(j))\rangle_e$. This suggests that the ETA model accurately captures the variance of behaviour within the life time of a single cell. We therefore do not consider switching between different 'modes' of movement [33,43,65]. However, we found that the inter-cellular variance $\sigma(\langle\tau(j)\rangle_t)$ is larger in experiment than predicted by the ETA model (inset Fig. 1c), indicating that the cell population may exhibit phenotypic heterogeneity.

In principle, a natural approach to tackle inter-cellular variability within the framework of our stochastic nonlinear model would be to infer single-cells models. For a population of *n* cells, we would then infer *n* models, and we could investigate how these would distribute over model space. However, due to the finite division time of the cells, there is fundamental bound on the information we can obtain per cell from the trajectories, which we find does not suffice to reliably infer the deterministic and stochastic components for the whole phase space. Instead, we therefore propose to rank the population of cells by their average dwell time $\langle\tau(j)\rangle_t$, and then split the population into subpopulations of cells with similar average hopping rates (Fig. 4a). Importantly, the ranking of the cells is uncorrelated with the trajectory length, suggesting that it is not determined by the cell cycle stage of the cell. We then infer separate stochastic nonlinear models for each of the subpopulations, generate trajectories for each model, and finally analyze a population of trajectories made up of these subpopulation-trajectories.

Interestingly, performing this procedure for only two subpopulations already significantly improves our estimate of the inter-cellular variance (reducing the relative error by more than a factor of 2), while not significantly affecting the prediction of other quantities such as the population mean and variance (Fig. 4c). However, further increasing the number of subpopulations does not appear to significantly improve predictions, suggesting that the dominant contribution to this heterogeneity can be captured by two phenotypes. These results indicate that CCV in this system is well approximated by a decomposition into two subpopulations.

As an alternative approach to subpopulation models, we tested a model in which we measured an overall rescaling prefactor for each single cell, in a similar spirit to our approach to the 'aging' effect. However, this approach does not predict the correct population averaged dwell time $\langle\tau\rangle_{t,e}$ and furthermore does not lead to an improvement in the estimate of the transition variance (Supplementary Section S3).

Given that a splitting into two subpopulations appears to be the simplest effective approach, we next asked whether there is an optimal way of doing so. To answer this question, we vary the fraction of cells included in the fast model. Interestingly, we find that the model performs optimally for a 50:50 split of the cell population (Fig. 4d). Thus, through the quantitative nature of our approach, we are able to identify the most natural way of decomposing the cell population into subpopulations, by optimizing the predictive power of the model. In general, one could imagine other scenarios than a predictive optimum at 50:50, for example, if there are two clearly distinct phenotypes in the population with an uneven distribution. In contrast, the actual variation of cellular identities in the experimental data appears to be continuous from very slow to very fast cells (Fig. 4a). This is further supported by the observation that the distributions of single cell observables, such as dwell times and speeds, exhibit unimodal distributions (Supplementary Section S2). The 50:50 decomposition into subpopulations therefore represents a first approximation to characterize the continuous variability in the ensemble, which we will thus use in the remainder of the paper.

To put the two-type decomposition to a test, we construct a model that accounts for both CCV and aging by using the slow and fast models, both including the same time-dependent prefactor $\alpha(t)$. Strikingly, this aging subpopulation model can accurately predict the mean-variance curve observed experimentally, indicating that both effects need to be accounted for to capture the behavioural variability of confined cell migration (Fig. 4e). As expected, the mean-variance curves of the subpopulations are thus also accurately predicted by the subpopulation models when aging is included (Supplementary Section S4). Furthermore, the aging subpopulation model accurately captures the other statistics that were already well-predicted by the ETA model, such as the dwell time, the position and velocity distribution and the velocity auto-correlation function (Supplementary Section S4). Taken together, these results indicate that the cell migration in this system exhibits phenotypic heterogeneity detectable at the level of cell trajectories that can be captured by an equation of motion with two sets of parameters, for a 'fast' and a 'slow' subpopulation of approximately equal sizes; on top of this, the subpopulations homogenously age over time, in the form of a gradual slowing down which is similar for all cells.

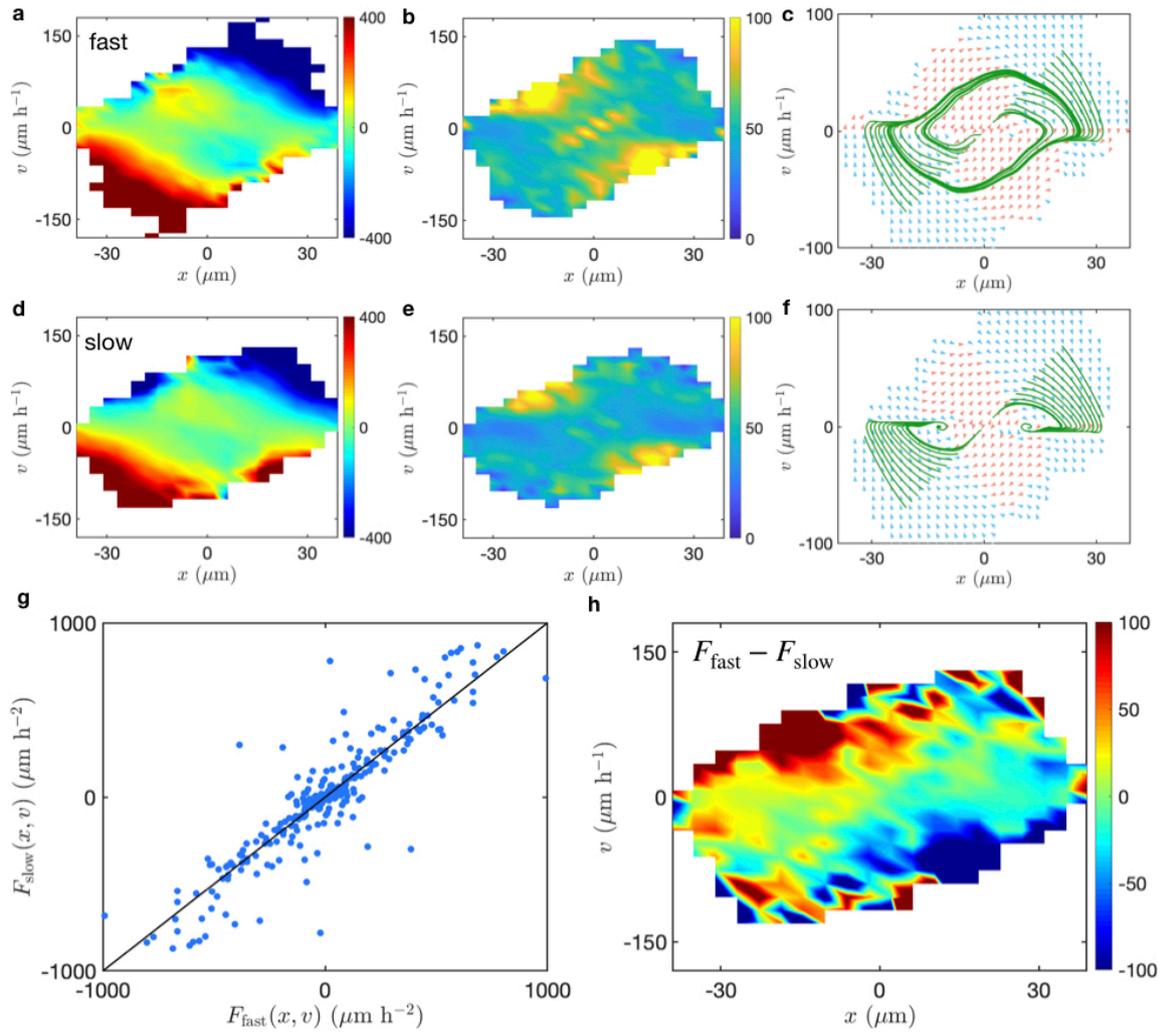

**Fig 5. Subpopulation models ($L$ = 16 μm).** (a), (d) Deterministic terms $F(x,v)$ in units of μm/h². (b), (e) The noise strengths $\sigma(x,v)$ in units of μm/h^(3/2). (c), (f) Deterministic flow fields and trajectories. (g) Local deterministic accelerations in the slow model plotted against those in the fast model. (h) The difference between the slow and fast deterministic terms as a function of position in phase space. Panels (a)-(c) correspond to the fast subpopulation, while panels (d)-(f) correspond to the slow subpopulation.

**Subpopulations exhibit distinct non-linear dynamics**

Our theoretical framework allows us to pinpoint the key aspects of the cellular dynamics that are sensitive to population heterogeneity. To conceptualize the deterministic dynamics, we plot the flow fields in $xv$-phase space that describe how the purely deterministic system evolves in time (Fig. 5c, f). Here, the region of phase space in which the cell is deterministically accelerating is colored in orange. This shows that the cells exhibit a deterministic amplification that drives the transitions across the constriction, as we observe deterministic acceleration just before the cell enters the thin constriction. Interestingly, we find that both subpopulation models exhibit similar qualitative features as the ETA model, including this characteristic amplification stage (Fig. 5a, c, d, f). Moreover, the dynamics exhibit similar noise hotspots as in the ETA model (Fig. 5b, e) in states where the cells initiate transition attempts. However, we find that the deterministic acceleration in the amplification phase is larger in the fast model, and similarly the noise hotspot is more pronounced in this case. Interestingly, the differences between the fast and slow models cannot be explained by an overall rescaling factor (Fig. 5g), but rather manifest as local deviations in the phase space, which are largest in the regions of phase space corresponding to states where the cells initiate the transition (Fig. 5h).

The two subpopulation models furthermore exhibit distinct non-linear dynamics: while the fast model exhibits a limit cycle, the slow model is bistable (Fig. 5c, f). These bistable dynamics are however highly excitable, as the amplification region extends nearly all the way to the stable fixed points, meaning that a small perturbation suffices for the system to perform a large deterministic excursion to the other fixed point. The transition from limit cycle and bistable dynamics is therefore the consequence of a relatively small shift in parameters, where only a subtle change in flow is required to modify the system from bistable to limit cycle dynamics. Interestingly, while the subpopulations exhibit distinct deterministic dynamics, the ETA model is dominated by the limit cycle behaviour of the fastest cells (Fig. 1d). This appears to be a robust result, as both the early- and late-stage models inferred separately for the first and second half of the experiment exhibit limit cycle dynamics, despite the slowing down of the average dynamics (Supplementary Section 1).

**Discussion**

In this work, we have developed a quantitative approach for detecting and characterizing temporal and inter-individual variability in the behaviour of migrating cells. To detect cell-to-cell variability (CCV), we used an ensemble-averaged stochastic equation of cell motion as a reference to determine the inherent variability of single cells migrating within a confining two-state micropattern. This ensemble-based approach successfully captures the key statistics of the cell trajectories, such as the dwell time distribution, the distributions of position and velocity and the velocity correlations. However, here we identify statistics aimed at quantifying the variability of behaviour between cells in the population, and show that they are underestimated by the ensemble-averaged model. We extend our framework to capture this variability, by performing a decomposition of the cell population into subpopulations of similarly behaving cells. This approach then captures the overall variability of the motion and reveals that the subpopulations exhibit distinct deterministic nonlinear dynamics. Taken together, our results show which statistics of the migration are sensitive to time- and population-heterogeneity, and which are well explained by an ensemble-averaged approach.

Using the variance of transition times as a marker for CCV, we found that the experimentally observed variability significantly exceeds that predicted by a single stochastic process. This indicates that migrating cells exhibit CCV that affects the behaviour at the level of cell trajectories. Previous studies have shown that the proteome of single human cells in a population exhibits a broad distribution [12], and this variability is frequently interpreted as a consequence of the inherent stochasticity of various intra-cellular processes [11]. However, CCV at the level of cell behaviour in a population context have in many cases been shown to be determined by environmental factors [17,64,66,67], such as external cues like local cell density, cell-cell contacts and relative location in a cell cluster. Moreover, previous measurements of the phenotypic variability of single cells on micropatterns have shown that the distributions of subcellular organelles are remarkably constant [68], a finding that has been interpreted to indicate that there is little intrinsic variability in isolated cells [64]. In contrast, here we show that isolated migrating cells in identical confining micro-environments do exhibit significant variability that is detectable at the behavioural level.

The CCV-sensitive predictions of our model are improved significantly by decomposing the cell population into subpopulations. Interestingly, evenly dividing cells into two subpopulations of slow and fast cells leads to a model that performs well in predicting intercellular variance. This decomposition reveals subtle differences in the system-level dynamics of the subpopulations: the deterministic dynamics of the fast cells exhibit a limit cycle, while those of the slow cells are an excitable bistable system. However, an inspection of our data does not indicate that there are two discrete types of cell behaviour, but rather suggest a continuum of variation to which a two-type decomposition is a reasonable coarse-grained approximation. At the single-cell level, we would thus expect a smooth cross-over from limit cycle to bistable behaviour. This distinction is similar to our previous finding that two different cell lines exhibit these two classes of non-linear dynamics [32]. Specifically, we found that the dynamics of the non-cancerous and less invasive [69,70] MCF10A cell line can be described as a bistable system, which is in contrast to the limit cycle behaviour of the cancerous MDA-MB-231 cell line that we consider here (Fig. 1d). The presence of these two distinct behaviours thus appears to be a more common feature of confined two-state cell migration. Our observation here that both behaviours are represented among cells of a single cell line highlights the breadth of phenotypic diversity present in this population.

Furthermore, we showed that cells gradually slow down over their life-time, an effect that can be captured by an overall rescaling factor of the equation of motion. The observed time-dependence could be related to the fact that cells grow in size over time [71,72] or could reflect a mechano-sensitive adaption of cells to their confining environment [73]. Similar 'aging' effects have been observed in cell populations [74], due to the gradual 'jamming' of the growing cell sheet. Our results show that there may be an additional contribution at the single-cell level to such observations. Together, the 'aging' effect and the two-type CCV implementation lead to a model that quantitatively captures the long time-scale transition distribution which characterizes the migratory behaviour of cells in the two-state micropattern.

In summary, our work provides a quantitative assessment of the diversity of behavioural phenotypes in a cell population. The unique geometry of the two-state micropattern allows a natural definition of variability in terms of the variances of the transition behaviour, which can be compared to a reference theory for this system. Our quantitative framework may help our understanding of cell-to-cell variability in collective systems [21–24,64,74] and provide useful tools for inverse approaches that attempt to correlate heterogeneity in behaviour to differences in the proteomes of single cells [64,75–78].

## Methods

**Experiments**
We use microscale plasma-induced protein patterning [79] to create standardized cell micro-environments for MDA-MB-231 human breast carcinoma epithelial cells (DSMZ). The cells are cultured in Minimum Essential Medium (MEM, c.c. pro), which is supplemented with 10% FBS (Gibco) and 2mM L-Glutamine (c.c. pro). Cells are grown at 37°C in an atmosphere with 5% $CO_2$. For passaging, cells are washed and trypsinized for 3 min. This cell solution is centrifuged for 3 min at 1000 rcf, and then cells are re-suspended in MEM. Approximately 10,000 cells are added per μ-dish (ibidi GmbH) and left to adhere for 4h in the incubator. Subsequently, the medium is exchanged to L-15 medium containing L-glutamine (Gibco, supplemented with 10% FCS) and 25 nM Hoechst 33342 (Invitrogen) for fluorescent staining of cell nuclei. Time-lapse measurements are performed in on an IMIC digital microscope (TILL Photonics) or on a Nikon Eclipse Ti microscope. Images in brightfield and fluorescent mode are acquired every 10 min for up to 50 h. The samples kept at 37°C in a heated chamber (ibidi GmbH or Okolab) throughout the measurements. To extract cell trajectories, first a band pass filter is applied to the images of the nuclei. Then images are binarized and *ImageJ*'s Analyze Particles plugin [80] is used to determine the center of the nuclei. A reference boundary of the micropattern is extracted from the brightfield images to yield absolute cell positions. Further details can be found in ref. [32].

**Inference of the equation of motion**
The position of the cell nucleus $x(t)$ is recorded at a time interval $\Delta t$ = 10 min in the experiment. Velocity and acceleration are directly calculated as numerical derivatives, given by $v(t) = (x(t) - x(t - \Delta t))/\Delta t$ and $a(t) = (v(t + \Delta t) - v(t))/\Delta t$, respectively. We interpret equation (1) in the Itô-sense throughout, and infer $F(x,v) = \langle \dot{v}|x,v\rangle$ and $\sigma^2(x,v) = \Delta t \langle [\dot{v} - F(x,v)]^2|x,v\rangle$ by conditional averaging [58,59,81] in a coarse-grained grid of 30x30 bins. To make optimal use of the available data, we exploit the inversion symmetry of the micropattern, which renders the dynamical terms (anti-)symmetric: $F(x,v) = -F(-x,-v)$ and $\sigma(x,v) = \sigma(-x,-v)$. More details can be found in [32].

## Acknowledgements

We thank C. Schreiber, D. S. Fischer, E. Frey, J. M. Lober, J. Messelink, G. Gradziuk, M. Schmitt, N. Arlt, P. Ronceray, and G. Stephens and his group for inspiring discussions. This work was supported by grants from the German Science Foundation (DFG) through the Collaborative Research Center (SFB) 1032 (Projects B01 and B12). D.B.B. is supported by a DFG fellowship within the Graduate School of Quantitative Biosciences Munich (QBM) and by the Joachim Herz Stiftung.

## References

1. Spudich JL, Koshland DE. 1976 Non-genetic individuality: chance in the single cell. *Nature* **262**, 467–471.
2. Aldridge BB, Fernandez-suarez M, Heller D, Ambravaneswaran V, Irimia D, Toner M, Fortune SM. 2012 Growth and Antibiotic Susceptibility. *Science (80-. ).* **335**, 100–104.
3. Umehara S, Inoue I, Wakamoto Y, Yasuda K. 2007 Origin of individuality of two daughter cells during the division process examined by the simultaneous measurement of growth and swimming property using an on-chip single-cell cultivation system. *Biophys. J.* **93**, 1061–1067. (doi:10.1529/biophysj.106.098061)
4. Min TL, Mears PJ, Chubiz LM, Rao C V., Golding I, Chemla YR. 2009 High-resolution, long-term characterization of bacterial motility using optical tweezers. *Nat. Methods* **6**, 831–835. (doi:10.1038/nmeth.1380)
5. Jordan D, Kuehn S, Katifori E, Leibler S. 2013 Behavioural diversity in microbes and low-dimensional phenotypic spaces. *Proc. Natl. Acad. Sci.* **110**, 14018–14023. (doi:10.1073/pnas.1308282110)
6. Li L, Cox EC, Flyvbjerg H. 2011 'Dicty dynamics': Dictyostelium motility as persistent random motion. *Phys. Biol.* **8**, 046006. (doi:10.1088/1478-3975/8/4/046006)
7. Berman GJ, Choi DM, Bialek W, Shaevitz JW. 2014 Mapping the stereotyped behaviour of freely moving fruit flies. *J. R. Soc. Interface* **11**, 20140672–20140672. (doi:10.1098/rsif.2014.0672)
8. Forkosh O, Karamihalev S, Roeh S, Engel M, Alon U, Nussbaumer M, Flachskamm C, Kaplick P, Shemesh Y. 2018 Identity domains in complex behaviour: Toward a biology of personality. (doi:http://dx.doi.org/10.1101/395111)
9. Eysenck HJ. 1953 *The structure of human personality.* New York, NY, US: Methuen.
10. McCrae RR, Costa Jr. PT. 2003 Personality in adulthood: A five-factor theory perspective, 2nd ed. *Personal. adulthood A five-factor theory Perspect. 2nd ed.* (doi:10.4324/9780203428412)
11. Niepel M, Spencer SL, Sorger PK. 2009 Non-genetic cell-to-cell variability and the consequences for pharmacology. *Curr. Opin. Chem. Biol.* **13**, 556–561. (doi:10.1016/j.cbpa.2009.09.015)
12. Sigal A *et al.* 2006 Variability and memory of protein levels in human cells. *Nature* **444**, 28–31. (doi:10.1038/nature05316)
13. Raj A, van Oudenaarden A. 2008 Nature, Nurture, or Chance: Stochastic Gene Expression and Its Consequences. *Cell* **135**, 216–226. (doi:10.1016/j.cell.2008.09.050)
14. Wagner A, Regev A, Yosef N. 2016 Revealing the vectors of cellular identity with single-cell genomics. *Nat. Biotechnol.* **34**, 1145–1160. (doi:10.1038/nbt.3711)

15. Altschuler SJ, Wu LF. 2010 Cellular Heterogeneity: Do Differences Make a Difference? *Cell* **141**, 559–563. (doi:10.1016/j.cell.2010.04.033)
16. Holland AJ, Cleveland DW. 2008 Beyond Genetics: Surprising Determinants of Cell Fate in Antitumor Drugs. *Cancer Cell*. (doi:10.1016/j.ccr.2008.07.010)
17. Cohen AA *et al.* 2008 Dynamic Proteomics of Individual Cancer Cells in Response to a Drug. *Science (80-. ).* **322**, 1511–1516. (doi:10.1126/science.1160165)
18. Feinerman O, Veiga J, Dorfman JR, Germain RN, Altan-Bonnet G. 2008 Variability and Robustness in T Cell Activation from Regulated Heterogeneity in Protein Levels. *Science (80-. ).* **321**, 1081–1084. (doi:10.1126/science.1158013)
19. Gascoigne KE, Taylor SS. 2008 Cancer Cells Display Profound Intra- and Interline Variation following Prolonged Exposure to Antimitotic Drugs. *Cancer Cell* **14**, 111–122. (doi:10.1016/j.ccr.2008.07.002)
20. Wieser S, Weghuber J, Sams M, Stockinger H, Schütz GJ. 2009 Cell-to-cell variability in the diffusion constants of the plasma membrane proteins CD59 and CD147. *Soft Matter* **5**, 3287–3294. (doi:10.1039/B902266J)
21. Camley BA, Rappel W. 2017 Cell-to-cell variation sets a tissue-rheology – dependent bound on collective gradient sensing. *Proc. Natl. Acad. Sci. USA* **114**, E10074–E10082. (doi:10.1073/pnas.1712309114)
22. Li X, Das A, Bi D. 2019 Mechanical heterogeneity in tissues promotes rigidity and controls cellular invasion.
23. Copenhagen K, Malet-engra G, Yu W, Scita G, Gov N, Gopinathan A. 2018 Frustration-induced phases in migrating cell clusters. , 1–10.
24. Lee RM, Yue H, Rappel WJ, Losert W. 2017 Inferring single-cell behaviour from largescale epithelial sheet migration patterns. *J. R. Soc. Interface* **14**. (doi:10.1098/rsif.2017.0147)
25. Petrie RJ, Doyle AD, Yamada KM. 2009 Random versus directionally persistent cell migration. *Nat. Rev. Mol. Cell Biol.* **10**, 538–49. (doi:10.1038/nrm2729)
26. Danuser G, Allard J, Mogilner A. 2013 Mathematical Modeling of Eukaryotic Cell Migration: Insights Beyond Experiments. *Annu Rev Cell Dev Biol.* **29**, 501–528. (doi:10.1146/annurev-cellbio-101512-122308.Mathematical)
27. Gail MH, Boone CW. 1970 The Locomotion of Mouse Fibroblasts in Tissue Culture. *Biophys. J.* **10**, 980–993. (doi:10.1016/S0006-3495(70)86347-0)
28. Wu P-H, Giri A, Sun SX, Wirtz D. 2014 Three-dimensional cell migration does not follow a random walk. *Proc. Natl. Acad. Sci. U. S. A.* **111**, 3949–54. (doi:10.1073/pnas.1318967111)
29. Selmeczi D, Mosler S, Hagedorn PH, Larsen NB, Flyvbjerg H. 2005 Cell motility as persistent random motion: theories from experiments. *Biophys. J.* **89**, 912–931. (doi:10.1529/biophysj.105.061150)
30. Selmeczi D, Li L, Pedersen LII, Nrrelykke SF, Hagedorn PH, Mosler S, Larsen NB, Cox EC, Flyvbjerg H. 2008 Cell motility as random motion: A review. *Eur. Phys. J. Spec. Top.* **157**, 1–15. (doi:10.1140/epjst/e2008-00626-x)
31. Takagi H, Sato MJ, Yanagida T, Ueda M. 2008 Functional analysis of spontaneous cell movement under different physiological conditions. *PLoS One* **3**, e2648. (doi:10.1371/journal.pone.0002648)
32. Brückner DB, Fink A, Schreiber C, Röttgermann PJF, Rädler JO, Broedersz CP. 2019 Stochastic nonlinear dynamics of confined cell migration in two-state systems. *Nat. Phys.* **15**, 595–601. (doi:10.1038/s41567-019-0445-4)

33. Potdar AA, Jeon J, Weaver AM, Quaranta V, Cummings PT. 2010 Human mammary epithelial cells exhibit a bimodal correlated random walk pattern. *PLoS One* **5**, e9636. (doi:10.1371/journal.pone.0009636)
34. Vestergaard CL, Pedersen JN, Mortensen KI, Flyvbjerg H. 2015 Estimation of motility parameters from trajectory data: A condensate of our recent results. *Eur. Phys. J. Spec. Top.* **224**, 1151–1168. (doi:10.1140/epjst/e2015-02452-5)
35. Thüroff F, Goychuk A, Reiter M, Frey E. 2019 Bridging the gap between single cell migration and collective dynamics. (doi:http://dx.doi.org/10.1101/548677)
36. Dietrich M, Le Roy H, Brückner DB, Engelke H, Zantl R, Rädler JO, Broedersz CP. 2018 Guiding 3D cell migration in deformed synthetic hydrogel microstructures. *Soft Matter* (doi:10.1039/C8SM00018B)
37. Albert PJ, Schwarz US. 2016 Dynamics of Cell Ensembles on Adhesive Micropatterns: Bridging the Gap between Single Cell Spreading and Collective Cell Migration. *PLoS Comput. Biol.* **12**, 1–34. (doi:10.1371/journal.pcbi.1004863)
38. Segerer FJ, Thüroff F, Piera Alberola A, Frey E, Rädler JO. 2015 Emergence and persistence of collective cell migration on small circular micropatterns. *Phys. Rev. Lett.* **114**, 228102. (doi:10.1103/PhysRevLett.114.228102)
39. Ziebert F, Swaminathan S, Aranson IS. 2012 Model for self-polarization and motility of keratocyte fragments. *J. R. Soc. Interface* **9**, 1084–1092. (doi:10.1098/rsif.2011.0433)
40. Camley BA, Zhang Y, Zhao Y, Li B, Ben-Jacob E, Levine H, Rappel W-J. 2014 Polarity mechanisms such as contact inhibition of locomotion regulate persistent rotational motion of mammalian cells on micropatterns. *Proc. Natl. Acad. Sci.* **111**, 14770–14775. (doi:10.1073/pnas.1414498111)
41. Goychuk A, Brückner DB, Holle AW, Spatz JP, Broedersz CP, Frey E. 2018 Morphology and Motility of Cells on Soft Substrates.
42. Albert PJ, Schwarz US. 2016 Modeling cell shape and dynamics on micropatterns. *Cell Adhes. Migr.* **10**, 516–528. (doi:10.1080/19336918.2016.1148864)
43. Metzner C, Mark C, Steinwachs J, Lautscham L, Stadler F, Fabry B. 2015 Superstatistical analysis and modelling of heterogeneous random walks. *Nat. Commun.* **6**, 7516. (doi:10.1038/ncomms8516)
44. Green BJ, Panagiotakopoulou M, Pramotton FM, Stefopoulos G, Kelley SO, Poulikakos D, Ferrari A. 2018 Pore shape defines paths of metastatic cell migration. *Nano Lett.* **18**, 2140–2147.
45. Paul CD, Mistriotis P, Konstantopoulos K. 2017 Cancer cell motility: Lessons from migration in confined spaces. *Nat. Rev. Cancer* **17**, 131–140. (doi:10.1038/nrc.2016.123)
46. Wolf K *et al.* 2013 Physical limits of cell migration: Control by ECM space and nuclear deformation and tuning by proteolysis and traction force. *J. Cell Biol.* **201**, 1069–1084. (doi:10.1083/jcb.201210152)
47. Paul CD, Shea DJ, Mahoney MR, Chai A, Laney V, Hung WC, Konstantopoulos K. 2016 Interplay of the physical microenvironment, contact guidance, and intracellular signaling in cell decision making. *FASEB J.* **30**, 2161–2170. (doi:10.1096/fj.201500199R)
48. Wilson K, Lewalle A, Fritzsche M, Thorogate R, Duke T, Charras G. 2013 Mechanisms of leading edge protrusion in interstitial migration. *Nat. Commun.* **4**, 1–12. (doi:10.1038/ncomms3896)
49. Liu YJ, Le Berre M, Lautenschlaeger F, Maiuri P, Callan-Jones A, Heuzé M, Takaki T,

Voituriez R, Piel M. 2015 Confinement and low adhesion induce fast amoeboid migration of slow mesenchymal cells. *Cell* **160**, 659–672. (doi:10.1016/j.cell.2015.01.007)
50. Charras G, Sahai E. 2014 Physical influences of the extracellular environment on cell migration. *Nat. Rev. Mol. Cell Biol.* **15**, 813–824. (doi:10.1038/nrm3897)
51. Wolf K, Alexander S, Schacht V, Coussens LM, von Andrian UH, van Rheenen J, Deryugina E, Friedl P. 2009 Collagen-based cell migration models in vitro and in vivo. *Semin. Cell Dev. Biol.* **20**, 931–941. (doi:https://doi.org/10.1016/j.semcdb.2009.08.005)
52. Friedl P, Alexander S. 2011 Cancer invasion and the microenvironment: Plasticity and reciprocity. *Cell* **147**, 992–1009. (doi:10.1016/j.cell.2011.11.016)
53. Gritsenko PG, Ilina O, Friedl P. 2012 Interstitial guidance of cancer invasion. *J. Pathol.* **226**, 185–199. (doi:10.1002/path.3031)
54. Davidson PM, Battistella A, Déjardin T, Betz T, Plastino J, Cadot B, Borghi N, Sykes C. 2019 Actin accumulates nesprin-2 at the front of the nucleus during confined cell migration. (doi:https://doi.org/10.1101/713982)
55. Renkawitz J *et al.* 2019 Nuclear positioning facilitates amoeboid migration along the path of least resistance. *Nature* **568**, 546–550. (doi:10.1038/s41586-019-1087-5)
56. Patteson AE, Pogoda K, Byfield FJ, Charrier EE, Peter A, Deptu P, Bucki R, Janmey PA. 2019 Loss of vimentin intermediate filaments decreases peri-nuclear stiffness and enhances cell motility through confined spaces.
57. Siegert S, Friedrich R, Peinke J. 1998 Analysis of data sets of stochastic systems. *Phys. Lett. Sect. A Gen. At. Solid State Phys.* **243**, 275–280. (doi:10.1016/S0375-9601(98)00283-7)
58. Ragwitz M, Kantz H. 2001 Indispensable Finite Time Corrections for Fokker-Planck Equations from Time Series Data. *Phys. Rev. Lett.* **87**, 254501. (doi:10.1103/PhysRevLett.87.254501)
59. Stephens GJ, Johnson-Kerner B, Bialek W, Ryu WS. 2008 Dimensionality and Dynamics in the Behaviour of C. elegans. *PLoS Comput Biol* **4**, e1000028. (doi:10.1371/journal.pcbi.1000028)
60. Svoboda K, Mitra PP, Block SM. 1994 Fluctuation analysis of motor protein movement and single enzyme kinetics. *Proc. Natl. Acad. Sci. USA* **91**, 11782–11786. (doi:10.1073/pnas.91.25.11782)
61. Taniguchi Y, Choi PJ, Li G-W, Chen H, Babu M, Hearn J, Emili A, Xie XS. 2010 Quantifying E. coli Proteome and Transcriptome with Single-Molecule Sensitivity in Single Cells. *Science (80-. ).* **329**, 533–538. (doi:10.1126/science.1188308)
62. Sherman MS, Lorenz K, Lanier MH, Cohen BA. 2015 Cell-to-Cell Variability in the Propensity to Transcribe Explains Correlated Fluctuations in Gene Expression. *Cell Syst.* **1**, 315–325. (doi:10.1016/j.cels.2015.10.011)
63. Ben-David U *et al.* 2018 Genetic and transcriptional evolution alters cancer cell line drug response. *Nature* **560**, 325–330. (doi:10.1038/s41586-018-0409-3)
64. Snijder B, Pelkmans L. 2011 Origins of regulated cell-to-cell variability. *Nat. Rev. Mol. Cell Biol.* **12**, 119–125. (doi:10.1038/nrm3044)
65. Schreiber C, Segerer FJ, Wagner E, Roidl A, Rädler JO. 2016 Ring-Shaped Microlanes and Chemical Barriers as a Platform for Probing Single-Cell Migration. *Sci. Rep.* **6**, 26858. (doi:10.1038/srep26858)
66. Snijder B, Sacher R, Rämö P, Damm E-M, Liberali P, Pelkmans L. 2009 Population

context determines cell-to-cell variability in endocytosis and virus infection. *Nature* **461**, 520.
67. Colman-Lerner A, Gordon A, Serra E, Chin T, Resnekov O, Endy D, Gustavo Pesce C, Brent R. 2005 Regulated cell-to-cell variation in a cell-fate decision system. *Nature* **437**, 699–706. (doi:10.1038/nature03998)
68. Schauer K, Duong T, Bleakley K, Bardin S, Bornens M, Goud B. 2010 Probabilistic density maps to study global endomembrane organization. *Nat. Methods* **7**, 560–566. (doi:10.1038/nmeth.1462)
69. Mak M, Reinhart-King CA, Erickson D. 2011 Microfabricated physical spatial gradients for investigating cell migration and invasion dynamics. *PLoS One* **6**, e20825. (doi:10.1371/journal.pone.0020825)
70. Kraning-Rush CM, Carey SP, Lampi MC, Reinhart-King CA. 2013 Microfabricated collagen tracks facilitate single cell metastatic invasion in 3D. *Integr. Biol.* **5**, 606–616. (doi:10.1039/c3ib20196a)
71. Théry M. 2010 Micropatterning as a tool to decipher cell morphogenesis and functions. *J. Cell Sci.* **123**, 4201–4213. (doi:10.1242/jcs.075150)
72. Suffoletto K, Ye N, Meng F, Verma D, Hua SZ. 2015 Intracellular forces during guided cell growth on micropatterns using FRET measurement. *J. Biomech.* **48**, 627–635. (doi:10.1016/j.jbiomech.2014.12.051)
73. Gegenfurtner FA, Jahn B, Wagner H, Ziegenhain C, Enard W, Geistlinger L, Rädler JO, Vollmar AM, Zahler S. 2018 Micropatterning as a tool to identify regulatory triggers and kinetics of actin-mediated endothelial mechanosensing. *J. Cell Sci.* **131**, jcs212886. (doi:10.1242/jcs.212886)
74. Garcia S, Hannezo E, Elgeti J, Joanny J-F, Silberzan P, Gov NS. 2015 Physics of active jamming during collective cellular motion in a monolayer. *Proc. Natl. Acad. Sci.* **112**, 15314–15319. (doi:10.1073/pnas.1510973112)
75. Chan TE, Stumpf MPH, Babtie AC. 2017 Gene Regulatory Network Inference from Single-Cell Data Using Multivariate Information Measures. *Cell Syst.* **5**, 251–267. (doi:10.1016/j.cels.2017.08.014)
76. Todorov H, Cannoodt R, Saelens W, Saeys Y. 2019 Network Inference from Single-Cell Transcriptomic Data BT - Gene Regulatory Networks: Methods and Protocols. In (eds G Sanguinetti, VA Huynh-Thu), pp. 235–249. New York, NY: Springer New York. (doi:10.1007/978-1-4939-8882-2_10)
77. Li B, You L. 2013 Predictive power of cell-to-cell variability. *Quant. Biol.* **1**, 131–139. (doi:10.1007/s40484-013-0013-3)
78. Sachs K, Perez O, Pe'er D, Lauffenburger DA, Nolan GP. 2005 Causal Protein-Signaling Networks Derived from Multiparameter Single-Cell Data. *Science (80-. ).* **308**, 523–529. (doi:10.1126/science.1105809)
79. Segerer FJ, Röttgermann PJF, Schuster S, Piera Alberola A, Zahler S, Rädler JO. 2016 Versatile method to generate multiple types of micropatterns. *Biointerphases* **11**, 011005. (doi:10.1116/1.4940703)
80. Schneider CA, Rasband WS, Eliceiri KW. 2012 NIH Image to ImageJ: 25 years of image analysis. *Nat. Methods* **9**, 671–675. (doi:10.1038/nmeth.2089)
81. Siegert S, Friedrich R, Peinke J. 1998 Analysis of data sets of stochastic systems. *Phys. Lett. A* **243**, 275–280. (doi:10.1016/S0375-9601(98)00283-7)